\RequirePackage{ifpdf}
\ifpdf 
\documentclass[pdftex]{sigma}
\else
\documentclass{sigma}
\fi

\numberwithin{equation}{section}

\newtheorem{thm}{Theorem}[section]
\newtheorem{prop}{Proposition}[section]

\begin{document}

\allowdisplaybreaks

\renewcommand{\PaperNumber}{081}

\FirstPageHeading

\ShortArticleName{On a Recently Introduced Fifth-Order Bi-Hamiltonian Equation}

\ArticleName{On a Recently Introduced Fifth-Order\\ Bi-Hamiltonian Equation\\  and Trivially Related Hamiltonian Operators}

\Author{Daryoush TALATI and Refik TURHAN}

\AuthorNameForHeading{D.~Talati and R.~Turhan}

\Address{Department of Engineering Physics, Ankara University 06100 Tando\u{g}an Ankara, Turkey}
\Email{\href{mailto:talati@eng.ankara.edu.tr}{talati@eng.ankara.edu.tr}, \href{mailto:turhan@eng.ankara.edu.tr}{turhan@eng.ankara.edu.tr}}

\ArticleDates{Received April 25, 2011, in f\/inal form August 18, 2011;  Published online August 20, 2011}

\Abstract{We show  that a recently introduced f\/ifth-order bi-Hamiltonian equation with a~dif\/ferentially constrained arbitrary function by A.~de Sole, V.G.~Kac and M.~Wakimoto is not a new one but a higher symmetry of a third-order equation. We give an exhaustive list of cases of the arbitrary function in this equation, in each of which the associated equation is inequivalent to the equations in the remaining cases. The equations in each of the cases are linked to equations known in the literature by invertible transformations. It is shown that the new Hamiltonian operator of order seven, using which the introduced equation is obtained, is trivially related to a known pair of f\/ifth-order and third-order compatible Hamiltonian operators. Using the so-called trivial compositions of
lower-order Hamiltonian operators, we give nonlocal generalizations of some higher-order Hamiltonian operators.}

\Keywords{bi-Hamiltonian structure; Hamiltonian operators}

\Classification{37K05; 37K10}

\section{Introduction}

A hierarchy of evolution equations $u_{t_{i}}=F_{i}[u]$ are called bi-Hamiltonian integrable if by two compatible Hamiltonian operators
(HO's) $J$ and $K$ there exist a Magri scheme
\begin{gather*}
u_{t_{i}}=F_{i}[u]=K\delta_{u}\int h_{i}[u]{\mathrm d}x=J\delta_{u}\int h_{i+1}[u]{\mathrm d}x,\qquad i=0,1,2,3,\dots
\end{gather*}
with conserved densities $h_{i}[u]$ which are assumed to exist for all~$i$, where $\delta_{u}$ is the variational derivative. HO's are skew-adjoint operators satisfying Jacobi identity. Two HO's are said to be compatible if their arbitrary linear combinations are also HO. A direct byproduct of a bi-Hamiltonian structure determined by compatible pair of HO's $K$ and formally invertible $J$ is the recursion operator $R=KJ^{-1}$ which maps an equation  $F_{i}$ to the next equation $F_{i+1}=RF_{i}$ in the symmetry hierarchy $u_{t_{i}}=F_{i}$, $i=0,1,2,\dots$. Square brackets, like~$F[u]$ denote dif\/ferential functions of~$x$, $u$ and $x$-derivatives of $u$ up to some f\/inite order.

Because of their central role in the integrability theory, various operator classes are classif\/ied for their HO content. Local and scalar HO's of order~1 and~3 are classif\/ied by Gel'fand--Dorfman, Astashov--Vinogradov, Mokhov, Olver in~\cite{GD,AV,Mok1,Olv1}, and the 5th-order ones in~\cite{Cook1} by Cooke. Recently in \cite{SKW}, A.~de Sole, V.G.~Kac and M.~Wakimoto (deSKW) classif\/ied  HO's of order 7 up to 13
under equivalence up to contact transformations. They also gave a conjecture for HO's of order grater than 13.

Contact transformations, i.e.\ the invertible transformations $x=P(y,v,v')$ and $u=Q(y,v,v')$ leaving the equation $w=0$ of the contact form $w=\mathrm{d}u-u'\mathrm{d}x$ invariant, provide a natural class of equivalence transformations for HO's because they preserve locality of HO's and, moreover, the order of the transformed HO remains equal to that of the original one~\cite{Mok1}. In what follows the HO's and the other relevant objects like symmetries, conserved densities etc. which can be transformed into each other by contact transformations will be called {\em equivalent}.

Here we consider an integrable 5th-order equation with an arbitrary function subject to a~dif\/ferential constraint, obtained by a pair of compatible newly obtained  7th-order and 3rd-order HO's in~\cite{SKW} which is announced to be new. We show  that the obtained equation is not a new one but a higher symmetry of a 3rd-order equation which is equivalent to one of three particular equations given in Mikhailov--Shabat--Sokolov (MSS) classif\/ication~\cite{MSS} according to the arbitrary function it contains. We give a lower-order recursion operator and show that the newly obtained 7th-order HO is trivially related to a pair of known 5th-order and 3rd-order HO's. By using successive trivial compositions of the compatible 5th- and 3rd-order HO's we provide  some nonlocal generalizations of the HO's of order grater than 7 obtained in deSKW classif\/ication.

\section{The f\/ifth-order equation}

The result of the deSKW classif\/ication of 7th-order HO's in \cite{SKW}  is the following theorem.
\begin{thm}[\cite{SKW}]\label{ThSKW}
Any HO of order $7$ is equivalent  either to a quasiconstant
coefficient skew-adjoint differential operator or to the operator $H_{(7,c(x))}+b^{2}D^{3}$, where
\[
 H_{(7,c(x))}=-B^{*}_{(3,c(x))}\cdot D \cdot B_{(3,c(x))}, \qquad B_{(3,c(x))}=\frac{1}{u}D\frac{1}{u}D^{2}+c(x)D-\frac{1}{2}c^{\prime}(x),
\]
and $c^{\prime \prime \prime}(x)=0$, $b={\rm const}$. These two types of HO's are not equivalent. The HO's $H_{(7,c(x))}+b^{2}D^{3}$
and $H_{(7,c_{1}(x))}+b_{1}^{2}D^{3}$ are equivalent if and only if $\alpha^{2}c_{1}(x)=c(\alpha^{3}x+ \beta)$ and $\alpha^{2} b_{1}=\pm b$ for
some constants $\alpha \ne 0$ and $\beta$. Such a HO is equivalent to a linear combination of the operators $H^{(j,0)}$ if
and only if $c(x)=c={\rm const}$, and one has  $H_{(7,c)}=H^{(7,0)}+2cH^{(5,0)}+c^{2}H^{(3,0)}$.
\end{thm}

In the above theorem, {\em quasiconstant} refers to arbitrary functions of $x$ only and $(*)$ denotes (formal) adjoint. $H^{(N,0)}$ are HO's def\/ined for $N=2n+3 \ge 3$ by
\begin{gather}
 H^{(N,0)}=D^{2}\cdot \left(\frac{1}{u}D \right)^{2n}\cdot D, \qquad D^{n}=\frac{\mathrm{d}^{n}\phantom{x}}{\mathrm{d}x^{n}}. \label{HDham}
\end{gather}

Using the compatible pair of HO's $H_{(7,c(x))}$ with $c^{\prime \prime \prime}(x)=0$ and $D^{3}$, the following equation
\begin{gather}
u_{t_{1}} = H_{(7,c(x))}\delta_{u}\int h_{0}\mathrm{d}x
  = \left( \frac{u^{(4)}}{u^{5}}-15\frac{u'u'''}{u^{6}}-10\frac{u''^{2}}{u^{6}} +105\frac{u''u'^{2}}{u^{7}}-105\frac{u'^{4}}{u^{8}}
\right.\nonumber\\
\phantom{u_{t_{1}} =}{}
+\left(2c(x)-\frac{1}{4}c''(x)x^{2}+\frac{1}{2}c'(x)x\right)\frac{u''}{u^{3}}-\left(\frac{3}{2}c'(x)x+6c(x)-\frac{3}{4}c''(x)x^{2}\right)\frac{u'^{2}}{u^{4}}
\nonumber\\
\phantom{u_{t_{1}} =}{}
+5c'(x)\frac{u'}{u^{3}} -\frac{5}{4}\frac{c''(x)}{u^{2}}-\frac{15}{16}c(x)^{2}+\frac{9}{16}c(x)c''(x)x^{2} -\frac{9}{8}c(x)c'(x)x
\nonumber\\
\left.\phantom{u_{t_{1}} =}{}
 +\frac{3}{64}c''(x)^{2}x^{4}-\frac{3}{16}c'(x)c''(x)x^{3}+\frac{3}{16}c'(x)^{2}x^{2} \right)'=D^{3}\delta_{u}\int h_{1}\mathrm{d}x
 \label{NE}
\end{gather}
is introduced as the f\/irst member of a new integrable symmetry hierarchy.
The associated initial conserved density which is a Casimir functional for the HO $D^{3}$, is
\[
 h_{0}=-\frac{1}{2}x^{2}u,
\]
and the second conserved density is given as
\[
h_{1}=a(x)u+\frac{1}{u}\left(c(x)-\frac{1}{8}c''(x)x^{2}+\frac{1}{4}c'(x)x\right)-\frac{u'^{2}}{2u^{5}},
\]
with
\begin{gather*}
 a(x)=\frac{x^{2}}{32}\big({-}18c(x)^{2}-6c(x)c''(x)x^{2}+20c(x)c'(x)x\\
\phantom{a(x)=}{} -c''(x)^{2}x^{4}+5c'(x)c''(x)x^{3}-7c'(x)^{2}x^{2}\big).
\end{gather*}

Discarding the trivial symmetry $u_{t_{-1}}=0=D^{3}\delta_{u}\int h_{0} \mathrm{d}x$, the hierarchy constructed by the HO's $H_{(7,c(x))}$ with $c'''(x)=0$ and $D^{3}$ starts from the 5th-order equation~(\ref{NE}). The next symmetry $u_{t_{3}}=H_{(7,c(x))}\delta_{u}\int h_{1} \mathrm{d}x$ is of order~9.  The fourth-order recursion operator
\begin{gather}
R_{(4,c(x))}=H_{(7,c(x))}D^{-3} \label{rec4}
\end{gather}
gives the symmetries each being 4 orders higher than the one it succeeds.

\section{Lower-order symmetry and recursion operator}

Interestingly, for none of the functions $c'''(x)=0$, the 5th-order equation~(\ref{NE}) is equivalent to any of the 5th-order equations given in MSS classif\/ication~\cite{MSS} where the scalar equations of order up to 5 are extensively classif\/ied with respect to existence of suf\/f\/iciently many higher conserved densities for the existence of a formal symmetry.

From the recursion operator point of view, the order of the recursion operator $R_{(4,c(x))}$ in~(\ref{rec4}) does not match those of the 5th-order equations of Sawada--Kotera and Kaup--Kupershmidt both of which have recursion operators of order 6. Note that the order of a recursion operator is preserved by plenty of transformations.

All these facts lead us to reconsider the equation (\ref{NE}) for its lower-order symmetries at f\/irst. The result is the following proposition.

\begin{prop}
The $5$th-order equation~\eqref{NE} with $c'''(x)=0$ is a higher symmetry of the $3$rd-order equation
\begin{gather}
u_{t_{0}}=\frac{u'''}{u^{3}}-9\frac{u''u'}{u^{4}}+12\frac{u'^{3}}{u^{5}}-\frac{3}{2}c'(x)= \left(u^{-3}u''-3u^{-4}u'^{2}-\frac{3}{2}c(x)\right)'. \label{ord3}
\end{gather}
\end{prop}

If we further search for symmetries of order higher than 5 we obtain a 7th-order symmetry before the one at order 9. All these intermediate symmetries suggest existence of a 2nd-order recursion operator.  Indeed, it can be shown (but not needed in view of the results of next section) that

\begin{prop}The $2$nd-order operator
\begin{gather}
R_{(2,c(x))}=\left(H_{(7,c(x))}D^{-3}\right)^{\frac{1}{2}}=D^{2}\frac{1}{u}D\frac{1}{u}D^{-1}+c(x)+\frac{3}{2}c'(x)D^{-1} \label{rec2}
\end{gather}
is a hereditary recursion operator for equation~\eqref{ord3}.
\end{prop}

Since the inverse of $D$ is not uniquely def\/ined, by applying $R_{(2,c(x))}$ on the r.h.s.\ of (\ref{ord3}) we obtain the following linear combination
\begin{gather}
u_{t}=F_{1}+kF_{0},\label{lincomb}
\end{gather}
where
\begin{gather}
F_{1} = \left(\frac{u^{(4)}}{u^{5}}-15\frac{u'u'''}{u^{6}}-10\frac{u''^{2}}{u^{6}} +105\frac{u''u'^{2}}{u^{7}}-105\frac{u'^{4}}{u^{8}}\right.
\nonumber\\
\left.
\phantom{F_{1} =}{}
+\frac{5}{2}c(x)\frac{u''}{u^{3}}-\frac{15}{2}c(x)\frac{u'^{2}}{u^{4}}+5c'(x)\frac{u'}{u^{3}} - \frac{5}{4}\frac{c''(x)}{u^{2}}-\frac{15}{8}c(x)^{2} \right)'
\label{F1}
\end{gather}
and $F_{0}$ is the r.h.s.\ of equation~(\ref{ord3}). If the arbitrary constant $k$ in (\ref{lincomb}) is chosen to be the constant
\begin{gather}
k=-\frac{1}{4}c''(x)x^{2}+\frac{1}{2}c'(x)x-\frac{1}{2}c(x),\qquad c'''(x)=0, \label{sabit}
\end{gather}
then  the linear combination of symmetries (\ref{lincomb}) becomes (upon taking all the derivatives) exactly the equation~(\ref{NE}). So, the $F_{0}$ part
in equation~(\ref{NE}) is redundant. Rather than equation~(\ref{NE}) (with or without the $F_{0}$ part in it) we can concentrate on  its
3rd-order symmetry (\ref{ord3}) since it is a~lower-order symmetry in the hierarchy which is not of Lie-point type.

Having stepped down to order 3, in the following proposition, we divide the functions $c(x)$ which are constrained to satisfy $c'''(x)=0$, into three subclasses. For each subclass,  equations~(\ref{ord3}) are mutually inequivalent under contact transformations. We further relate equations from each subclass with the ones  given in literature.

\begin{prop}\label{uclu}
The equation~\eqref{ord3} for
\begin{itemize}\itemsep=0pt
\item[$i)$]$c(x)=c_{1}(x)$ where $c_{1}^{\prime}(x)=0$, is a special case of the equation~{\rm (4.1.34)} in {\rm \cite{MSS}}, whose potentiation is a symmetry in a Riemann hierarchy {\rm \cite{ON}}, which by a further hodograph transformation becomes a linear equation with only a $3$rd-order term;
\item[$ii)$]$c(x)=c_{2}(x)$ where $c_{2}^{\prime \prime}(x)=0$ and $c_{2}^{\prime}(x) \ne 0$, is equivalent to
\begin{gather}
u_{t}= \left( u^{-3}u''-3u^{-4}u'^{2} -3x \right)' \label{ord3b1}
\end{gather}
which is the equation~{\rm (4.1.23)} in {\rm \cite{MSS}}, through
\[
x \mapsto \left(\frac{2}{c_{2}'(x)}\right)^{\frac{1}{3}}x,\qquad t \mapsto \left(\frac{2}{c_{2}'(x)}\right) t;
\]
\item[$iii)$]$c(x)=c_{3}(x)$ where $c_{3}^{\prime \prime \prime}(x)=0$ and $c_{3}^{\prime \prime}(x) \ne 0$, is equivalent to
\begin{gather}
u_{t}= \left( u^{-3}u''-3u^{-4}u'^{2} +\frac{3}{2}x^{2} \right)' \label{ord3b}
\end{gather}
which is the equation~{\rm (4.1.24)} in {\rm \cite{MSS}}, through
\[
x \mapsto \left(\frac{-2}{c_{3}''(x)}\right)^{\frac{1}{4}}x-\frac{c_{3}'(0)}{c_{3}''(x)},\qquad t \mapsto \left(\frac{-2}{c_{3}''(x)}\right)^{\frac{3}{4}} t.
\]
\end{itemize}
\end{prop}

Let us note that under the contact transformations, each of the cases in the above proposition is an equivalence class.
And the three cases exhaust all functions $c(x)$ such that $c'''(x)=0$. These cases are the cases by which the HO's of form $H_{(7,c(x))}$ in Theorem \ref{ThSKW} are exhaustively divided into three equivalence class of HO's too.

\section{Lower-order Hamiltonian operator}

The recursion operator $R_{(4,c(x))}$ is a consequence of the compatible HO's $H_{(7,c(x))}$ with $c'''(x)=0$ and $D^{3}$ as the ratio given in (\ref{rec4}). Taking its square root we obtained the 2nd-order recursion operator $R_{(2,c(x))}$. Also considering the restored missing members of the symmetry hierarchy as above, it is natural to ask whether $R_{(2,c(x))}$ has a bi-Hamiltonian factorization too. Upon assuming the second HO to be again $D^{3}$, the answer reads
\[
R_{(2,c(x))}=H_{(5,c(x))}D^{-3},
\]
where
\begin{gather}
H_{(5,c(x))}=D^{2}\frac{1}{u}D\frac{1}{u}D^{2}+c(x)D^{3}+\frac{3}{2}c'(x)D^{2}, \qquad c'''(x)=0, \label{h5c}
\end{gather}
is the HO given in Remark~3.8 in \cite{SKW} (denoted by $H_{(5,0,c(x))}$ there). It is equivalent to a HO obtained by Cooke in \cite{Cook1}.

So, as a summary the hierarchy obtained by $H_{(7,c(x))}$ with $c'''(x)=0$ and $D^{3}$ is nothing but a subset of the hierarchy that the compatible pair of HO's $H_{(5,c(x))}$ with $c'''(x)=0$ and $D^{3}$ gives rise to. Let us note, for the sake of completeness that the
3th-order equation (\ref{ord3}) is obtained by
\begin{gather*}
u_{t}=\left(u^{-3}u''-3u^{-4}u'^{2}-\frac{3}{2}c(x)\right)' = H_{(5,c(x))}\delta_{u}\int \left(-\frac{x^{2}}{2}u\right) \mathrm{d}x  \\
\hphantom{u_{t}=\left(u^{-3}u''-3u^{-4}u'^{2}-\frac{3}{2}c(x)\right)'}{}
 = D^{3}\delta_{u}\int \left(\frac{1}{2u}-\left(\frac{3}{2}D^{-2}(c(x)) + f(x)\right)u\right)\mathrm{d}x
\end{gather*}
with any function $f(x)$ such that $f'''(x)=0$. The next symmetry in the hierarchy
\[
u_{t}=H_{(5,c(x))}\delta_{u} \int \left(\frac{1}{2u}-\left(\frac{3}{2}D^{-2}(c(x))+f(x)\right)u\right)\mathrm{d}x,
\]
is either the equation~(\ref{F1}) for the choice of the constant $f''(x)=0$, or the linear combination~(\ref{NE}) for the choice $f''(x)=k$ where $k$ is the constant given in (\ref{sabit}).

\section{Trivially related Hamiltonian operators}

A HO  $K_{1}$ which is obtainable from other compatible HO's $K_{0}$ and $J$ as $K_{1}=K_{0}J^{-1}K_{0}$, is called {\em trivially related} in \cite{ON} since, as seen here, the pair $(K_{1},J)$ gives a subset of the structure of symmetries/conserved quantities that the pair $(K_{0},J)$ gives. The sequence of operators $K_{n}$ obtained by trivially composing compatible HO's $K_{0}$ and  $J$ as $K_{n}=(K_{0}J^{-1})^{n}K_{0}$, $n=1,2,\dots$ are proved to be HO in \cite{FF}. It can be further shown by constructing trivial compositions
\[
\left(J+\sum_{n=0}^{m}\lambda_{n}\left(K_{0}J^{-1}\right)^{n}K_{0}\right)J^{-1}\left(J+\sum_{n=0}^{m}\bar \lambda_{n}\left(K_{0}J^{-1}\right)^{n}K_{0}\right),\qquad \left(K_{0}J^{-1}\right)^{0}=1,
\]
of the successive partial sums $m=0,1,2,\dots$ of linear combinations $J+\sum\limits_{n=0}^{m}\lambda_{n}(K_{0}J^{-1})^{n}K_{0}$ with arbitrary constants
$\lambda_{n}$, $\bar \lambda_{n}$
and by induction on $m$ that all $K_{n}$, $n=0,1,2,\dots$ are mutually compatible HO's if so is the HO's $K_{0}$ and (formally) invertible~$J$.

As an example, consider the HO's $H^{(N,0)}$ and $H^{(M,0)}$ with  $N=2n+3 \ge 3$, $M=2m+3 \ge 3$, i.e.\ $n,m=0,1,2,3,\dots$,  of type~(\ref{HDham}) whose Darboux form was obtained in~\cite{Vodova}. Their trivial compositions{\samepage
\begin{gather*}
 H^{(2N-M,0)}=H^{(N,0)}\big(H^{(M,0)}\big)^{-1}H^{(N,0)}=D^{2}\cdot \left(\frac{1}{u}D \right)^{2s}\cdot D,
\end{gather*}
where $s=2n-m=0,\pm 1,\pm 2, \pm 3, \dots$ are all mutually compatible HO's.}

Trivially related HO's were f\/irst singled out in~\cite{ON} where the considered equations were of so-called hydrodynamic type. Equations of this type may possess plenty of Hamiltonian representations with more than two compatible HO's.
The 1st-order scalar equations in general~\cite{Nutku}, and the Riemann equation as a particular representative thereof~\cite{ON}, is an extreme case in the number of compatible Hamiltonian formulations admitted. They possess inf\/initely many Hamiltonian structures~\cite{FP} on the same set of variables not only with 1st-order HO's, even if the trivially related HO's are isolated \cite{RT}.

In deSKW classif\/ication \cite{SKW}, all the obtained higher-order HO's as well as the 5th-order ones obtained by Cooke in \cite{Cook1} are given in a particular symmetric form which is not only beautiful but also very suitable to observe possible trivial lower-order decompositions.
Taking into account that
\[
D^{2}B_{(3,c(x))}=H_{(5,c(x))}, \qquad c'''(x)=0,
\]
where the operator $B_{(3,c(x))}$ is given in Theorem~\ref{ThSKW} and $H_{(5,c(x))}$ in (\ref{h5c}), we observe
\[
H_{(7,c(x))}=H_{(5,c(x))}D^{-3}H_{(5,c(x))},
\]
which is a trivial composition of the HO's $H_{(5,c(x))}$ with $c'''(x)=0$ and $D^{3}$. A pair of which is known to be compatible from the classif\/ication of 5th-order HO's.

The fact that $H_{(7,c(x))}$ is trivially related to the HO's $H_{(5,c(x))}$ and $D^{3}$ has the following consequence:
\begin{prop}
The operator of order $7$
\[
J_{(7,c(x))}=H_{(7,c(x))}+aH_{(5,c(x))}+bD^{3},
\]
where $H_{(7,c(x))}$ is given in Theorem~{\rm \ref{ThSKW}} and $H_{(5,c(x))}$ in \eqref{h5c} with arbitrary constants $a$ and $b$ is HO iff $c'''(x)=0$.
\end{prop}

Note that the HO $H_{(7,c(x))}$, being a trivial composition of HO's $H_{(5,c(x))}$ and $D^{3}$, forms a~compatible triple with the latter two. In general, compatibility of one of the HO's, say $J_{2}$ of a compatible pair $(J_{1},J_{2})$ with a third one $J_{3}$ (not proportional to $J_{1}$ or $J_{2}$) does not imply compatibility of the pair  $(J_{1},J_{3})$. But this property holds automatically for trivially related HO's.

The HO $H_{(7,c(x))}$ is a strict trivial composition of the the pair $H_{(5,c(x))}$ and $D^{3}$ in the following sense:
\begin{prop} Let the functions $c(x)$ and $f(x)$ and a constant $a$ be such that $c'''(x)=0$, $f'''(x)=0$ and $a \ne 0$. Then
\[
H_{(7,f(x))}+aH_{(5,c(x))}+bD^{3}
\]
is HO only if $f(x)=c(x)$, for any value of constant $b$.
\end{prop}

Let us proceed with the compatible pair $H_{(7,c(x))}$ and $H_{(5,c(x))}$ with $c'''(x)=0$  to obtain the HO of order 9
\begin{gather*}
J_{(9,c(x))} = H_{(7,c(x))}\left(H_{(5,c(x))}\right)^{-1}H_{(7,c(x))}=\left(H_{(5,c(x))}D^{-3}\right)^{2}H_{(5,c(x))} \\
\phantom{J_{(9,c(x))}}{}
= B^{*}_{(3,c(x))}\left(D\frac{1}{u}D\frac{1}{u}D+c(x)D+\frac{1}{2}c'(x)-\frac{1}{2}c''(x)D^{-1}\right)B_{(3,c(x))},\qquad c'''(x)=0,
\end{gather*}
where $B_{(3,c(x))}$ is given in Theorem~\ref{ThSKW}. For $c''(x) \ne 0$ the HO  $J_{(9,c(x))}$ is an operator with a~nonlocal term $D^{-1}$ and thus outside of the deSKW classif\/ication.

As it was the case with $H_{(7,c(x))}$, $J_{(9,c(x))}$ cannot  give any new structure of symmetries or conservation laws other than the ones that the pair $H_{(5,c(x))}$ with $c'''(x)=0$ and $D^{3}$ gives because eventually it is a trivial composition of the latter two which implies the following:
\begin{prop}\label{ord9}
The operator of order $9$
\[
J_{(9,c(x))}+k_{1}H_{(7,c(x))}+k_{2}H_{(5,c(x))}+k_{3}D^{3}
\]
with arbitrary constants $k_{1}$, $k_{2}$ and $k_{3}$ is HO iff $c'''(x)=0$.
\end{prop}

The  $c''(x) = 0$ special case of Proposition~\ref{ord9} gives a local HO which agrees with the Theorem~3.11 of deSKW classif\/ication of HO's of order 9 in \cite{SKW}.

At order 11 there is the following trivial composition
\begin{gather*}
J_{(11,c(x))}\! = H_{(9,c(x))}\left(H_{(7,c(x))}\right)^{-1}\!H_{(9,c(x))}\!=\left(H_{(5,c(x))}D^{-3}\right)^{3}\!H_{(5,c(x))}\!
=-P^{*}_{(11,c(x))}DP_{(11,c(x))} ,\\
P_{(11,c(x))} = \left(\frac{1}{u}D\frac{1}{u}D+c(x)-\frac{1}{2}c'(x)D^{-1}\right)\!\left(\frac{1}{u}D\frac{1}{u}D^{2}+c(x)D-\frac{1}{2}c^{\prime}(x) \right), \qquad c'''(x)=0,
\end{gather*}
which is a local operator only for the case $c'(x){=}0$. Trivial compositions $\left(H_{(5,c(x))}D^{-3}\right)^{n}H_{(5,c(x))}$, $n=4,5,\dots$, i.e.\ at orders 13 and higher are HO's with nonlocal tail unless $c'(x)=0$.

Trivial compositions in the opposite direction $\left(D^{3}(H_{(5,c(x))})^{-1}\right)^{m}D^{3}$, $m=1,2,3,\dots$ are pseudodif\/ferential operators of order 1 and lower which are intractable. Therefore, trivial compositions of the local pair $(H_{(5,c(x))},D^{3})$ shows us conversely that in those cases with intractable HO pairs, there may exist an associated tractable, i.e.\ at least weakly nonlocal~\cite{Sergyeyev} or better local, pair of HO's which are trivial (de)compositions of the original intractable ones.

\section{Discussion}

We have shown that a recently introduced 5th-order integrable equation (\ref{NE}) is a linear combination of a 5th-order equation with its 3rd-order symmetry and related through the recursion operator $R_{(2,c(x))}$ in (\ref{rec2}), to the 3rd-order equation~(\ref{ord3}) as a higher symmetry. The equation~(\ref{ord3}), depending on the form of arbitrary function $c(x)$ in it, is equivalent to three well known equations. In other words, equation~(\ref{NE}) is a higher symmetry of equation~(\ref{ord3}) which is, in a sense, a~compact representation of three well known equations in a single expression.

The fact that none of the equations~(\ref{NE}) is in the MSS list of 5th-order integrable equations is understandable since, as noted, those 5th-order equations which are higher symmetries of lower-order equations  are omitted in the MSS list of 5th-order equations. Correspondence of each case in Proposition~\ref{uclu} with only one equation in MSS list is in agreement with the fact that the cases in Proposition~\ref{uclu} and the equations in the MSS classif\/ication are both divided according to the equivalence under contact transformations.

Integrable third-order scalar evolution equations were extensively classif\/ied using various def\/initions of integrability, see e.g.~\cite{MSS,GK} and the references therein.
As also noted in \cite{MSS}, equations~(\ref{ord3b1}) and  (\ref{ord3b}) are related to the KdV and pKdV equations respectively via potentiation and hodograph transformation which is not uniquely invertible and thus does not belong to the class of contact transformations. But if the class of contact transformations is extended, the ultimate 3rd-order equation is the KdV equation from which all the equations considered here can be derived~\cite{Sergei}.

On the side of HO's, we have explained by using trivial (de)compositions of HO's that the  5th-order equation (\ref{NE}) obtained by the compatible pair $(H_{(7,c(x))},D^{3})$ in deSKW classif\/ication is a higher symmetry in a f\/iner hierarchy constructed by the compatible pair $(H_{(5,c(x))},D^{3})$, a~trivial composition of which is the HO  $H_{(7,c(x))}$. Moreover, as a consequence of using trivial compositions of lower-order HO's we have obtained nonlocal generalizations of some of the higher-order HO's obtained in deSKW classif\/ication.

\subsection*{Acknowledgements}
R.T.\ thanks  Professors M.~G\"{u}rses and   A.~Karasu for stimulating discussions; and Professor V.G.~Kac for comments and clarif\/ication of the parametric nature of their results, removing a~misinterpretation with incorrect consequences. D.T.\ is supported by T{\"U}B{\.I}TAK PhD Fellowship for Foreign Citizens.

\pdfbookmark[1]{References}{ref}
\LastPageEnding

\end{document}